\documentclass[%
 aip,
 jmp,
 bmf,
 sd,
 rsi,
 amsmath,amssymb,
reprint,%
]{revtex4-2}

\usepackage{graphicx}
\usepackage{dcolumn}
\usepackage{bm}
\newcommand{\bea}{\begin{eqnarray}}
\newcommand{\eea}{\end{eqnarray}}
\newcommand{\be}{\begin{equation}}
\newcommand{\ee}{\end{equation}}

\newcommand{\bega}{\begin{align}}
\newcommand{\eega}{\end{align}}

\usepackage{subfigure}
\usepackage[export]{adjustbox}
\usepackage{amsmath,bm,amssymb}
\usepackage{hyperref}
\usepackage{euscript,array}
\usepackage{wrapfig}

\usepackage{epsfig}
\usepackage{amsthm}
\usepackage{graphics}
\usepackage{float}
\usepackage{graphicx} 

\usepackage{pstricks}
\usepackage{pst-eucl}
\usepackage{calc}
\usepackage{pst-3dplot}
\usepackage{pst-grad}
\usepackage{pst-plot,pst-math,pstricks-add}
\usepackage{pst-all}
\usepackage{pst-eps}

\usepackage[utf8]{inputenc}
\usepackage[T1]{fontenc}
\usepackage{mathptmx}

\begin{document}

\preprint{AIP/123-QED}

\title{One-loop effective action of the ${\mathbb C}P^{N-1}$ model at large $\mu\beta$}


\author{Antonino Flachi}
  \email{flachi@phys-h.keio.ac.jp}
 \affiliation{Department of Physics, \& Research and Education Center for Natural Sciences,\\ ~~~Keio University, Hiyoshi, Yokohama, Japan}
\author{Guglielmo Fucci}%
 \email{fuccig@ecu.edu}
\affiliation{ 
Department of Mathematics, East Carolina University, Greenville, NC 27858, USA
}%

\date{\today}

\begin{abstract}
In this note we consider a non-linear, large-$N$ ${\mathbb C}P^{N-1}$ sigma model on a finite size interval with periodic boundary conditions, at finite temperature and chemical potential in the regime of $\beta \mu$ large. Our goal is to extend previous calculations and obtain the coefficients of the derivative expansion of the one-loop effective action in the region of $\beta \mu$ large by carrying out the appropriate analytical continuation. This calculation complements previous results and allows us to conclude that the ground state remains homogeneous in this regime as long as it is assumed to be a slowly varying function of the spatial coordinates. While this is reasonable at the two extremes of small or large chemical potential, for intermediate values of the chemical potential and small enough temperature, one might expect (by analogy with other models) that lower energy crystalline solutions may exist. In this case a simple derivative expansion, like the one discussed here, would need to be modified in order to capture these features. 
\end{abstract}

\maketitle

\section{\label{sec:intro}Introduction}

Sigma models are of particular importance in the ambit of both condensed matter and high energy physics due to the fact that they can model many interesting physical phenomena. Perhaps the best known example is the $O(3)$ sigma model which is closely related to Heisenberg's model of anti-ferromagnetism. The $O(3)$ sigma model is used to describe interacting spins, where the interaction is introduced in the form of a constraint on the degrees of freedom of the field. The constraint is usually implemented by requiring the dynamical fields to be of fixed length. This condition allows the phases of the dynamical fields to be free to vary.   
In this model, field interactions arise because any change in one of the multiplet components must be accompanied by corresponding changes in the other components so that the length of the multiplet remains constant (as imposed by the constraint). It is possible to extend this model to include $N$ fields. This generalization is particularly important since it allows to study the model in the limit of large number of fields. More importantly, in this limit, the model can be solved exactly. A related classic example is the two-dimensional ${\mathbb C}P^{N-1}$ model, that is described by a $1+1$ dimensional field theory of $N$ complex scalar fields $n_i$ ($i=1,2,\cdots, N$) with an action of the form 
\bea
\EuScript S = \int dx dt \left| D_{\mu} n_i \right|^2,
\label{eqt1}
\eea 
and $D_\mu = \partial_\mu -i A_\mu$ with $A_\mu$ being a U(1) gauge field. The scalars $n_i$ obey the constraint,
\bea
\left| n_i \right|^{2} = r,
\label{eqt2}
\eea 
which represents a condition that causes interactions among the fields. The ${\mathbb C}P^{N-1}$ model has received considerable attention in view of its rich vacuum structure featuring asymptotic freedom, dynamical mass generation, and confinement. These characteristics make the ${\mathbb C}P^{N-1}$ model a useful toy model of QCD. A general introduction to the subject together with some of the original references about the ${\mathbb C}P^{N-1}$ model can be found, for example, in \cite{Zinn-Justin:2002,Shifman:2012,Polyakov:1975rr,Polyakov:1975yp,Bardeen:1976zh,Brezin:1976qa,DAdda:1978vbw,Witten:1978qu}.

Interest in this class of models (mainly $O(N)$ or ${\mathbb C}P^{N-1}$ models) has recently been revived. In particular, a number of works have been focused on analyzing the behavior of models such as $O(N)$ or ${\mathbb C}P^{N-1}$ in the presence of boundaries. A subset of these investigations can be found in references~\cite{Hong:1994,Shifman:2007rc,Gorsky:2013rpa,Monin:2015xwa,Milekhin2:2016fai,Bolognesi:2016zjp,Pikalov:2020lrb,Flachi:2017xat,Nitta3:2018azz,Pavshinkin:2017kwz,Betti:2017zcm,Bolognesi:2018njt,Bolognesi:2019rwq,Gorsky:2018lnd,Ishikawa:2019tnw,Ishikawa:2019oga,Flachi:2020pvn,Bonanno:2018xtd,Fujimori:2019skd,Berni:2019bch,Pelissetto:2019iic,Flachi:2019jus,Flachi:2019yci}.

One of the prominent issues being analyzed is related to the question of how the ground state responds to the presence of boundaries or, equivalently, to boundary conditions. Naturally, in the presence of topologically nontrivial (e.g., Dirichlet, Neumann, Robin) boundary conditions, the ground state, defined as the extremum of the effective action (it is important to notice that we do not define the ground state as the extremum of the effective action \textit{at zero temperature}, but simply as the extremum of the effective action), cannot be homogeneous. However, in the simpler case of periodic boundary conditions both homogeneous and inhomogeneous ground states are allowed. Usually, one would expect the homogeneous ground state to be energetically favorable. However, in the presence of external conditions, even in flat space, there are situations in which this does not hold true. Well-known examples of these unusual situations occur in quantum field theories with four-fermion interactions (e.g., the Gross-Neveu model and the Nambu-Jona Lasinio model are examples of the sort \cite{Nickel:2009ke}), where inhomogeneous ground states may become energetically favorable at intermediate values of the chemical potentials \cite{Nickel:2009ke}. Specifically, the ground state at high- or low-density and low-temperature is homogeneous, but it becomes inhomogeneous, with a resulting crystal structure, at intermediate densities.

In the present work, we will assume the field theory model (\ref{eqt1})-(\ref{eqt2}) to be defined on the interval $x \in \left[0,\ell\right]$. Then, our starting point is the following tree-level action  
\bea
\EuScript S = \int dx dt \left(\left| D_{\mu} n_i \right|^2 + M^2 \left(\left| n_i \right|^{2} - r\right)\right).
\label{eqtaux1}
\eea 
where the constraint $\left| n_i \right|^{2} = r$ is incorporated by means of a Lagrange multiplier $M^2$. In this note we focus on the case $A_\mu=0$, an assumption which has also been used in previous works on the subject\cite{Bolognesi:2016zjp,Bolognesi:2018njt,Bolognesi:2019rwq,Flachi:2019jus}. 

Finite temperature and chemical potential can be introduced in the above model and compete with the effect of boundary conditions imposed at the endpoints of the interval $\left[0,\ell\right]$. Notice that Green's functions satisfying spatial periodicity are analogous to thermal Green's functions, whose periodicity is instead set by the inverse temperature along the Wick rotated direction. This suggests that changes in the ground state may occur in the form of transitions between a low-temperature (or large-$\ell$), massive (Coulomb or confining) phase and a high-temperature (small-$\ell$) massless (Higgs or deconfining) phase, as a result of varying the size $\ell$ of the interval\cite{Monin:2015xwa}. At the same time, the Coleman-Mermin-Hoenberg-Wagner theory prevents the possibility of symmetry breaking and anticipates the locking of the ground state in a Coulomb phase, with no transition at all expected to occur \cite{Bolognesi:2019rwq}.

Rather than assume \textit{a priori} that no inhomogeneous ground state exists, it is preferable to entertain the possibility that the ground state may be inhomogeneous (away from non-periodic boundary conditions and in the presence of a chemical potential), and eventually discard that possibility \textit{a posteriori} by using the theory. To this end,  we need to assume a spatially-dependent ground state from the beginning of our analysis. A general approach to deal with this situation in the large-$N$ limit has been discussed, for example, in references \cite{Flachi:2019jus,Flachi:2019yci} where a derivative expansion for the one-loop effective action $\Gamma \left[M^2 \right]$ as a functional of the mass gap $M^2$ has been obtained in the form,
\bea
\Gamma \left[M^2 \right]&\sim& 
{\beta}\int_0^\ell d x 
\left\{ 
\alpha_0 + \alpha_1  M^2 + \alpha_2 M^4 + \alpha_3 \left(M^6 + \alpha_4 \left( \nabla \left(M^2\right)\right)^2 \right)
+ \cdots
\right\}.
\label{e1}
\eea
In the above expression, the coefficients $\alpha_p$ depend on the temperature and chemical potential, and the dots denote higher order terms as well as non-analytic contributions that may arise from infra-red and non-perturbative effects. Also, notice that the sign of the coefficients of the derivative terms determines whether spatially varying solutions increase or decrease the effective action.

We should remark here that the above expansion is nothing but the heat-kernel asymptotic expansion of the effective action for the problem under consideration. In general, this expansion is not convergent for any value of the temperature or chemical potential (or of other external fields which may be present). However, there exist specific ranges of values of the parameters involved in the expansion which make it an extremely good approximation suitable for a detailed analysis of the model (the typical ones considered in the literature are the high-temperature range or the weak-curvature range). In the present case, we will construct an expansion valid for $\mu\beta$ large, and the main point of the paper is precisely to compute the coefficients of the expansion in this range. Furthermore, in order to make our approach formally consistent, we need to assume that the contribution of higher derivatives to the effective action are increasingly smaller. This means that we are excluding possible solutions that are rapidly varying in space. While this appears to be a formal assumption, it is physically reasonable to expect that such rapidly varying solutions require more energy to exist.

The quantity $\alpha_0$ represents the vacuum energy contribution (see reference \cite{Flachi:2020pvn}). The explicit form of the coefficients $\alpha_p$, with $p\geq 1$ has been obtained in \cite{Flachi:2019jus} at finite temperature ($T=1/\beta$), and in \cite{Flachi:2019yci} in the presence of a finite chemical potential $\mu$ associated with the first component of the complex parameter $n_i$. In those papers the calculations were performed in the range $\beta \mu < 2\pi$ which led the authors to conclude that the inhomogeneous phases are energetically unfavorable. This outcome is not surprising, due to the fact that the chemical potential was assumed to be small. 

The goal of this note is to relax the condition $\beta \mu < 2\pi$ and extend the computation of the coefficients $\alpha_p$ with $p\geq 1$ to the case $\beta \mu$ large; this case includes a setup with large chemical potential $\mu$.

In the following, after briefly explain how to obtain a derivative expansion of the effective action at large-$N$, we will describe how to perform the analytic continuation of the coefficients $\alpha_p$ for $\beta \mu \gg 1$, extending the results of \cite{Flachi:2020pvn}. By using the explicit expression for coefficients $\alpha_p$ for $\beta \mu \gg 1$ we prove that the ground state remains homogeneous also in the range in which $\beta \mu$ is large.

\section{One-loop effective action and its derivative heat-kernel expansion}
\label{sec2}

In this section, we briefly summarize how to obtain the derivative expansion of the bulk one-loop effective action obtained in Refs.~\cite{Flachi:2019jus,Flachi:2019yci}. For brevity, we simply outline the main steps and refer the reader to the original references for additional details. The formal expression for the one-loop effective action for the model (\ref{eqtaux1}) is
\bea
\EuScript  S^{E}_{\mbox{\tiny{eff}}} &=& 
{\beta} \int d x 
\left\{ 
\left(\sigma'\right)^2  +  M^2 \left( \left| \sigma \right|^{2} - r
\right) {- \mu^2 \sigma^2}
\right\}  \nonumber\\
&&+ {(N-1)\over 2} \sum_\pm \mbox{Tr} \log \left(-{\partial_x^2} - {\partial_\tau^2} +  M^2 {- \mu^2\pm 2 \mu {\partial_\tau}}\right),~~~~
\label{eqt3.1}
\eea
and it is identical to that of \cite{Flachi:2019yci} in the present case. In the expression (\ref{eqt3.1}), we have chosen the background-field configuration to lie along the $k=1$ direction, i.e., $n_1 = \sigma,~n_{k} =0$ with $k>1$. In addition, following Ref.\cite{Bruckmann:2014sla}, we made the simplifying assumption requiring that the chemical potential $\mu$ be associated only with the first component of the complex parameter $n_i$. The above determinant can be analyzed in terms of the following zeta function
\bea
\zeta_\pm(s)= \sum_{k=0}^\infty\sum_{n=-\infty}^\infty \left(p^{(s)}_{k} + \left(2\pi n/\beta {\pm i \mu}\right)^2\right)^{-s},
\label{eqt4}
\eea
where the $p^{(s)}_{k}$ are the eigenvalues of the operator $\partial_x^2 + M^2$. 
According to the spectral zeta function regularization technique \cite{elizalde94,Avramidi}, the determinant in (\ref{eqt3.1}) can be evaluated in terms of the derivative at s=0 of the zeta function in (\ref{eqt4}). By using the Mellin transform and the heat-kernel asymptotic expansion (see Ref.~\cite{Flachi:2019jus} for details), Eqt.~(\ref{eqt4}) can be cast in the following form 
\bea
\zeta_\pm(s)&=& 
{1\over \Gamma(s)} {\beta \over {(4 \pi)^{(d+1+s)/2}}} 
\sum_{k=0}^\infty {\alpha}_{k} 
\int_{\Lambda^{-2}}^\infty {dt\over t^{(d+3-2k-s)/2}} \times \nonumber\\
&&\times \left(1 + 2 \sum_{n=1}^\infty \cosh \left(\beta \mu n\right) e^{-{\beta^2n^2 \over 4 t}}
\right),~~~~~~
\label{eqt15}
\eea
where $\Lambda$ is a UV regulator and
\bea
{\alpha}_0 &=&1,~~
{\alpha}_1 = - M^2,~~
{\alpha}_2 = {1\over 2} M^4 - {1\over 6} \Delta \left(\ell^2 M^2 \right)
, \nonumber\\
{\alpha}_3 &=& 
- {1\over 6}  M^6 + {1\over 12} \left( \nabla \left( M^2\right)\right)^2 + {1\over 6}  M^2  \Delta \left(M^2 \right)
- {1\over 60}  \Delta^2 \left(M^2 \right).
\nonumber
\label{eqt13}
\eea
By integrating (\ref{eqt15}) over $t$, then differentiating the resulting expression with respect to $s$, and by finally using the relation
\bea
\mbox{Tr} \log P=-\zeta'_{P}(0),
\eea
valid, in particular, for an elliptic and self-adjoint operator $P$,
one obtains the following expression for the bulk one-loop effective action
\bea
\EuScript  S^{E}_{\mbox{\tiny{eff}}} &=& 
{\beta}\int_0^\ell d x 
\left\{ 
\left(\nabla\sigma\right)^2  + M^2 \left( \left| \sigma \right|^{2} - r_\star
\right)-\mu^2\sigma^2 \right. \nonumber\\
&-&\left. {(N-1) \over {4 \pi}} 
\left[
-
\left(
\log \left({\beta^2\over\ell^2}\right)
- 2 \varpi'(1)
\right)M^2
+ {\beta^2\over 4} {}  \varpi'(3) M^4
\right.
\right.
\nonumber\\
&&
\left.
\left.
+ {\beta^4 \over 16} {} \varpi'(5) \left({1\over 6}  M^6 + {1\over 12} \left( \nabla \left(M^2\right)\right)^2 \right)
+\cdots
\right]
\right\}.
\label{eqt19}
\eea
In this expression the dots represent the Casimir energy of the system, the neglected higher orders terms and non-analytic contributions. The parameter $r_\star$ represents a renormalized coupling. The coefficients $\varpi_\pm\left(a\right) $ of the expansion are defined as
\bea
\varpi_\pm\left(a\right) =
\sum_{n=1}^\infty
\cosh \left(\pm z n\right) n^{-1+a},
\label{eqt16}
\eea
where $z=\beta \mu$. We would like to point out that the above results reproduce previously obtained results in the limit of vanishing chemical potential \cite{Flachi:2019jus}. So far, we have simply presented the results of Ref.~\cite{Flachi:2019jus} in which the coefficients $\varpi_\pm$ were evaluated by performing the analytic continuation in the region $0 \leq |z| \leq 2 \pi$. The goal of this work is to compute the coefficients $\varpi_\pm$ in the opposite limit; namely the one for large argument $z \gg 1$, i.e. $\beta \mu \gg 1$.\\

\begin{figure}
\begin{center}
\begin{tabular}{cc}
 \begin{pspicture*}(0.8,10.2)(9.25,13.5)
 \psline{->}(1.1,11)(9.2,11)
 \rput(8.8,10.8){$Re z$}
 \psline[ArrowInside=->,linestyle=dashed,linecolor=black,linewidth=1.4pt](1.1,11)(4.6,11)
 \psline[ArrowInside=->,linestyle=dashed,linecolor=black,linewidth=1.4pt](5.6,11)(9.,11)
 \psarc[ArrowInside=->,linestyle=dashed,linecolor=black,linewidth=1.4pt](5.1,11.){0.5}{0}{180}
 \psarc[ArrowInside=->,linestyle=dashed,linecolor=gray,linewidth=1.4pt](5.1,11.){0.5}{180}{0}
  \rput(5.6,11){$\bullet$}
  \rput(5.1,11){$\bullet$}
  \rput(4.6,11){$\bullet$}
 \rput(5.1,10.8){$x$}
 \rput(4.2,10.8){$x-\epsilon$}
 \rput(6.,10.8){$x+\epsilon$}
 \rput(5.5,11.6){$C_{\epsilon}^{+}$}
 \rput(5.5,10.4){$C_{\epsilon}^{-}$}
   \end{pspicture*}
\end{tabular}
\end{center}
\caption{Contour $\gamma$ (dashed line) used in formula (\ref{eq19}) showing both upper, $C_{\epsilon}^{+}$, and lower, $C_{\epsilon}^{-}$, branches.}
\label{figura_contorno}
\end{figure}
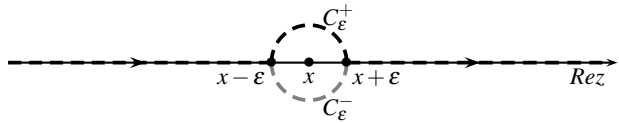

\section{Analytic continuation of the functions $\varpi_\pm$ for large argument.}

The analytic continuation of the function $\varpi_\pm$ for large values of the parameter $z$ can be obtained from the analytic continuation of the following function for $a\not\in \mathbb{N}_{+}$
\bea
f(a,x)&=&2\sum_{n=1}^{\infty}\cosh(x n)n^{a-1}=\textrm{Li}_{1-a}\left(e^{-x}\right)+\textrm{Li}_{1-a}\left(e^{x}\right),
\label{eqt17}
\eea
where $\textrm{Li}_{s}\left(w\right)$ denotes the polylogarithmic function. For $x\gg1$, the function $\textrm{Li}_{1-a}\left(e^{-x}\right)$ is exponentially small and, hence, will have a negligible contribution to the asymptotic expansion. The only relevant contributions to the large-$x$ asymptotic expansion of Eqt. (\ref{eqt17}) come from $\textrm{Li}_{1-a}\left(e^{x}\right)$.

In order to obtain the desired asymptotic expansion, we start with the Bose-Einstein integral representation of the polylogarithmic function 
\be
\textrm{Li}_{p}\left(e^{x}\right)=\frac{1}{\Gamma(p)}\int_{0}^{\infty}\frac{t^{p-1}}{e^{t-x}-1}dt\;,
\label{eq18}
\ee
which is valid for $x<0$ and in the semi-plane $\Re(p)>0$. This representation needs to be extended to positive values of the parameter $x$ to obtain a large-$x$ asymptotic expansion. However, when extending the integral representation to $x>0$, one needs to circumvent the pole of the integrand that occurs on the real axis at $t=x$. This pole can be avoided by considering, instead of the real integral in (\ref{eq18}), the following complex integral
\be
\int_{\gamma}\frac{z^{p-1}}{e^{z-x}-1}dz\;,
\label{eq19}
\ee
where the contour $\gamma$ (see Fig.\ref{figura_contorno}) consists of the segment $[0,x-\epsilon]$, with $\epsilon>0$, along the real line, followed by a semi-circle of radius $\epsilon$, $C^{\pm}_{\epsilon}$, (from above or below) around the pole at $z=x$ and ending with the infinite segment $[x+\epsilon,\infty]$. 
Along this contour one has
\begin{widetext}
\bea
\int_{\gamma}\frac{z^{p-1}}{e^{z-x}-1}dz&=&\int_{0}^{x-\epsilon}\frac{t^{p-1}}{e^{t-x}-1}dt
+\int_{C^{\pm}_{\epsilon}}\frac{z^{p-1}}{e^{z-x}-1}dz+\int_{x-\epsilon}^{\infty}\frac{t^{p-1}}{e^{t-x}-1}dt\;.
\label{eq20}
\eea
By using Cauchy's residue theorem to compute the integral over the semicircle and by subsequently taking the limit as $\epsilon\to 0$ one obtains
\bea
\textrm{Li}_{p}\left(e^{x}\right)=\mp i\pi \frac{x^{p-1}}{\Gamma(p)}+\frac{1}{\Gamma(p)}\left[-\int_{0}^{x}t^{p-1}dt-\int_{0}^{x}\frac{t^{p-1}}{e^{x-t}-1}dt+\int_{x}^{\infty}\frac{t^{p-1}}{e^{t-x}-1}dt\right]\;.
\label{eq21}
\eea
The first integral on the right-hand-side can be computed for $p>1$ and then analytically extended to $p\in\mathbb{C}$, to give
\bea
\textrm{Li}_{p}\left(e^{x}\right)=\mp i\pi \frac{x^{p-1}}{\Gamma(p)}-\frac{x^{p}}{\Gamma(p+1)}-\frac{1}{\Gamma(p)}\left[\int_{0}^{x}\frac{t^{p-1}}{e^{x-t}-1}dt-\int_{x}^{\infty}\frac{t^{p-1}}{e^{t-x}-1}dt\right]\;.
\label{eq22}
\eea
In Eqt. (\ref{eq22}), we set $y=x-t$ in the first integral and $y=t-x$ in the second integral. In this way one obtains 
\bea
\int_{0}^{x}\frac{(x-y)^{p-1}}{e^{y}-1}dy-\int_{0}^{\infty}\frac{(x+y)^{p-1}}{e^{y}-1}dy=\int_{0}^{\infty}\frac{(x-y)^{p-1}}{e^{y}-1}dy-\int_{x}^{\infty}\frac{(x-y)^{p-1}}{e^{y}-1}dy-\int_{0}^{\infty}\frac{(x+y)^{p-1}}{e^{y}-1}dy\;.\nonumber\\
\label{eq23}
\eea
By utilizing the integral representation (\ref{eq18}) it is not difficult to realize that 
\be
\int_{x}^{\infty}\frac{(x-y)^{p-1}}{e^{y}-1}dy=
(-1)^{p-1}\int_{0}^{\infty}\frac{t^{p-1}}{e^{t+x}-1}dt=(-1)^{p-1}\Gamma(p)\textrm{Li}_{p}\left(e^{-x}\right)\;,
\ee
which implies that the integral is exponentially small for $x\to\infty$ and can, hence, be ignored. The last remark allows us to write
\be
\int_{0}^{x}\frac{(x-y)^{p-1}}{e^{y}-1}dy-\int_{0}^{\infty}\frac{(x+y)^{p-1}}{e^{y}-1}dy\simeq \int_{0}^{\infty}\frac{(x-y)^{p-1}-(x+y)^{p-1}}{e^{y}-1}dy\;.
\label{eq24}
\ee
The integral appearing on the right-hand-side can be split into a sum of an integral over the interval $[0,x]$ and another one over the interval $[x,\infty]$. The integral over the semi-infinite portion, $[0,\infty]$ can be proved to be exponentially small. This implies that the relevant contributions to the integral in (\ref{eq24}) come from the region $[0,x]$ and since $y<x$ we can utilize the binomial expansion
\be
(x+y)^{p-1}-(x-1)^{p-1}=2x^{p}\sum_{k=1}^{\infty}\frac{\Gamma(p)y^{2k-1}x^{-2k}}{(2k-1)!\Gamma(p-2k+1)}\;.
\label{eq25}
\ee
The expansion (\ref{eq25}) allows us to obtain
\be
\int_{0}^{\infty}\frac{(x-y)^{p-1}-(x+y)^{p-1}}{e^{y}-1}dy=-2\sum_{k=1}^{\infty}\frac{\Gamma(p)x^{p-2k}}{(2k-1)!\Gamma(p-2k+1)}\int_{0}^{\infty}\frac{y^{2k-1}}{e^{y}-1}dy+O(e^{-x})\;.
\label{eq26}
\ee
\end{widetext}
By recalling the integral representation of the Riemann zeta function
\be
\zeta_{R}(s)=\frac{1}{\Gamma(s)}\int_{0}^{\infty}\frac{t^{s-1}}{e^{t}-1}dt\;,
\ee
valid for $\Re(s)>1$, one finally arrives at the result
\bea
\int_{0}^{\infty}\frac{(x-y)^{p-1}-(x+y)^{p-1}}{e^{y}-1}dy=-\sum_{k=1}^{\infty}\frac{2\zeta_{R}(2k)}{\Gamma(p-2k+1)}x^{p-2k}+O(e^{-x})\;.
\eea
The last relation, substituted in Eqt. (\ref{eq22}),
provides the large-$x$ asymptotic expansion of $\textrm{Li}_{p}\left(e^{x}\right)$ when $\Re(p)>0$
and, consequently, of $f(a,x)$, with $a\not\in \mathbb{N}$, as follows
\bea
f(a,x)&=&\mp i\pi \frac{x^{-a}}{\Gamma(1-a)}-\frac{x^{1-a}}{\Gamma(2-a)}+2\sum_{k=1}^{\infty}\frac{\zeta_{R}(2k)}{\Gamma(2-2k-a)}x^{1-a-2k}+O(e^{-x})\;.
\label{eq27}
\eea

When $a\in\mathbb{N}_{+}$ we have that 
\be
\textrm{Li}_{1-a}\left(e^{x}\right)=\textrm{Li}_{-n}\left(e^{x}\right)\;,
\ee
with $n\in\mathbb{N}_{0}$ and the polylogarithmic function is expressed in terms of elementary functions. For $n=0$ we find
\be
\textrm{Li}_{0}\left(e^{x}\right)=\frac{1}{1-e^{-x}}\;,
\ee
and, hence, as $x\to\infty$ one has the asymptotic behavior
\be
\textrm{Li}_{0}\left(e^{x}\right)=1+O(e^{-x})\;.
\ee

For $n\in\mathbb{N}_{+}$ one has
\be
\textrm{Li}_{n}\left(e^{x}\right)=\sum_{k=0}^{n}k!S(n+1,k+1)\left(\frac{e^{x}}{1-e^{x}}\right)^{k+1}\;,
\ee
where $S(n,k)$ represent the Stirling numbers of the second kind. From the above expression is not very difficult to find the following large-$x$ asymptotic behavior of $f(n-1,x)$
\be
f(n-1,x)=\sum_{k=0}^{n}k!(-1)^{k+1}S(n+1,k+1)+O(e^{-x})\;.
\ee

The results of this section allow us to write down the asymptotic expansion for $z\to\infty$ of the functions $\varpi_\pm\left(a\right)$. By using the definition (\ref{eqt16}) and the function $f(a,x)$ in (\ref{eqt17}) and by noticing that the hyperbolic cosine is an even function of its argument we find the large-$z$ expansions 
\bea
\varpi_\pm\left(a\right)&=&\mp i\pi \frac{z^{-a}}{2\Gamma(1-a)}-\frac{z^{1-a}}{2\Gamma(2-a)}+\sum_{k=1}^{\infty}\frac{\zeta_{R}(2k)}{\Gamma(2-2k-a)}z^{1-a-2k}+O(e^{-z})\;,
\label{eq28}
\eea
valid for $a\not\in \mathbb{N}$ and 
\be
\varpi_\pm\left(a\right)=\frac{1}{2}\sum_{k=0}^{a+1}k!(-1)^{k+1}S(a+2,k+1)+O(e^{-z})\;,
\label{eq29}
\ee
valid instead for $a\in\mathbb{N}_{+}$.

\section{Discussion}

The function of interest in our discussion is actually the sum $\varpi\left(a\right)=\varpi_{+}\left(a\right)+\varpi_{-}\left(a\right)$ rather than the individual functions $\varpi_{+}\left(a\right)$ and $\varpi_{-}\left(a\right)$. As it is clear from (\ref{eq28}), the asymptotic expansion of the functions $\varpi_{+}\left(a\right)$ and $\varpi_{-}\left(a\right)$ when $a\not\in \mathbb{N}$ acquires an imaginary part whose sign depends on the way the pole on the real axis is avoided. Since $\varpi\left(a\right)$ appears in the one-loop effective action, which for our system must be real, we need to choose a way of avoiding the singularity such that the resulting sum $\varpi_{+}\left(a\right)+\varpi_{-}\left(a\right)$ is represented, for large $x$, by a real quantity. The requirement that $\varpi\left(a\right)$ be real leads us to avoid the pole on the real axis for the construction of the asymptotic expansion of $\varpi_{+}\left(a\right)$ in the \emph{opposite way} we avoid the same pole for $\varpi_{-}\left(a\right)$. In other words, if we choose, in the derivation of the asymptotic expansion for $\varpi_{+}\left(a\right)$, to avoid the singularity by utilizing the semi-circle $C_{\epsilon}^{+}$, then we must choose to avoid the same singularity by using $C_{\epsilon}^{-}$ when constructing the asymptotic expansion for $\varpi_{-}\left(a\right)$. Obviously the opposite choice is also allowed. This same method was used in \cite{Flachi:2019yci} in the range $\beta\mu \ll 1$.
With this understanding, the large-$z$ asymptotic expansion of $\varpi\left(a\right)$ can be found to be
\be
\varpi\left(a\right)=-\frac{z^{1-a}}{\Gamma(2-a)}+2\sum_{k=1}^{\infty}\frac{\zeta_{R}(2k)}{\Gamma(2-2k-a)}z^{1-a-2k}+O(e^{-z})\;,
\ee
for $a\not\in \mathbb{N}$ and 
\be
\varpi\left(a\right)=\sum_{k=0}^{a+1}k!(-1)^{k+1}S(a+2,k+1)+O(e^{-z})\;,
\ee
when $a\in\mathbb{N}_{+}$.
The expressions above allow us to compute the derivatives $\varpi'\left(q\right)$ with $q=1,3,5, \dots$. We find:
\bea
\varpi'\left(a\right)&=& -{z^{1-a}\over \Gamma(2-a)}\left(\log\left(z\right) - \psi^{(0)}(z)\right)\nonumber\\
&-&2\sum_{k=1}^{\infty}
\frac{z^{1-a-2k} \zeta_{R}(2k)}{\Gamma(2-2k-a)}\left(
\log\left(z\right)
-
{\psi^{(0)}(2-2k-a)}
\right)+O(e^{-z})\;,\label{omegaprime}
\eea
where $\psi^{(0)}(z)$ is the polygamma function of order $0$.

In order to investigate whether the ground state of our system becomes inhomogeneous, it is sufficient to inspect the coefficient of the derivative terms that include derivatives of $M^2$. Truncating the expansion to sixth order (see equation (\ref{eqt19})), we see that the relevant coefficient is $\varpi'(5)$. By utilizing formula (\ref{omegaprime}) we obtain:
\bea
\varpi'(5) = -{6\over z^4} + {40\pi^2\over z^6} + {112\pi^4\over z^8} + \dots\;.
\eea
For large $z$ ($z\gg2\pi$), the sign of the coefficient $\varpi'(5)$ is determined by the dominant term in the above expansion, namely $-{6z^{-4}}$. Since the dominant term is negative, we can conclude that $\varpi'(5) <0$. This simple result allows us to ascertain the (homogeneous \textit{vs} inhomogeneous) nature of the ground state. In fact, according to the arguments of Ref.~\cite{Bolognesi:2019rwq}, the effective action should be maximized as a functional of the Lagrange multiplier $M^2$. Thus, the negativity of the coefficient $\varpi'(5)$ indicates that the derivative term decreases the effective action in the region of large $z$. 
Naively, one would conclude that the above remarks imply that the Lagrange multiplier, and consequently the ground state, remains homogeneous in this regime. However, a more rigorous conclusion is actually weaker and this has to do with the validity of our approximation. Our calculation uses a derivative expansion of the effective action that is valid when $\beta \times \mu$ is large and \textit{the ground state (i.e., the condensate and the Lagrange multiplier) are not rapidly varying functions of the spatial coordinates.} So our conclusions do not exclude the possibility that a different inhomogeneous crystalline ground state does not exist, but that this ground state is not likely to be a slowly varying function of the spatial coordinates.

The present approach should be complemented by one which allows for the computation of the effective action in a regime where the ground state is allowed to vary rapidly in space. One possibility is to use different methods to sum all the derivative terms and expand in powers of the condensate. Another, perhaps more useful method would be to proceed fully numerically. In the case of inhomogeneous ground states, these are both challenging problems that we hope to consider in the future.

\begin{acknowledgments}
The support of the Japanese Society for the Promotion of Science (Grant-in-Aid for Scientific Research KAKENHI Grant n. 18K03626) is gratefully acknowledged. 
\end{acknowledgments}


\end{document}